\documentclass[prl,nofootinbib,superscriptaddress,twocolumn]{revtex4-1}
\usepackage[a4paper,left=1.5cm,right=1.5cm,top=3cm,bottom=3cm]{geometry}

\usepackage{verbatim}
\usepackage{color}
\usepackage{amsfonts,amssymb,mathrsfs,amsmath}
\usepackage{framed}
\usepackage{mdframed}
\usepackage{latexsym}
\usepackage{graphicx}
\usepackage{datetime}
\newdateformat{mydate}{\THEDAY{ }\monthname[\THEMONTH]{ }\THEYEAR}

\allowdisplaybreaks

\usepackage{tikz}
\usepackage{color}
\usepackage{framed}
\usepackage{hyperref}
\hypersetup{colorlinks, citecolor=bluscuro, linkcolor=black, urlcolor=bluscuro}
\definecolor{rossos}{cmyk}{0,1,1,0.55}
\definecolor{bluscuro}{rgb}{0.15, 0.2, .85}
\definecolor{bluchiaro}{cmyk}{1,.3,0.,0.1}

\definecolor{ForestGreen}{rgb}{0.13, 0.55, 0.13}

\newcommand{\be}{\begin{equation}}
\newcommand{\ee}{\end{equation}}

\newcommand{\llp}{\left [}
\newcommand{\rrp}{\right ]}
\newcommand{\lp}{\left (}
\newcommand{\rp}{\right )}

\def\lsim{\mathrel{\rlap{\lower4pt\hbox{\hskip0.5pt$\sim$}}
    \raise1pt\hbox{$<$}}}         %less than or approx. symbol
\def\gsim{\mathrel{\rlap{\lower4pt\hbox{\hskip0.5pt$\sim$}}
    \raise1pt\hbox{$>$}}}         %greater than or approx. symbol

\newcommand{\arXiv}[2]{\href{http://arxiv.org/pdf/#1}{{\tt [#2/#1]}}}
\newcommand{\arXivold}[1]{\href{http://arxiv.org/pdf/#1}{{\tt [#1]}}}

\begin{document}

\title{The Formation Probability of Primordial Black Holes }

\author{Matteo Biagetti}
\address{Institute for Fundamental Physics of the Universe, Via Beirut 2, 34151 Trieste, Italy}
\address{SISSA - International School for Advanced Studies, Via Bonomea 265, 34136 Trieste, Italy}
\address{Istituto Nazionale di Astrofisica, Osservatorio Astronomico di Trieste, via Tiepolo 11, 34143 Trieste, Italy}
\address{Istituto Nazionale di Fisica Nucleare, Sezione di Trieste, via Valerio 2, 34127 Trieste, Italy}

\author{Valerio De Luca}
\address{D\'epartement de Physique Th\'eorique and Centre for Astroparticle Physics (CAP), Universit\'e de Gen\`eve, 24 quai E. Ansermet, CH-1211 Geneva, Switzerland}
\address{Dipartimento di Fisica, Sapienza Università 
di Roma, Piazzale Aldo Moro 5, 00185, Roma, Italy}

\author{Gabriele Franciolini}
\address{D\'epartement de Physique Th\'eorique and Centre for Astroparticle Physics (CAP), Universit\'e de Gen\`eve, 24 quai E. Ansermet, CH-1211 Geneva, Switzerland}

\author{Alex Kehagias}
\address{Physics Division, National Technical University of Athens, 15780 Zografou Campus, Athens, Greece}

\author{Antonio~Riotto}
\address{D\'epartement de Physique Th\'eorique and Centre for Astroparticle Physics (CAP), Universit\'e de Gen\`eve, 24 quai E. Ansermet, CH-1211 Geneva, Switzerland}

\date{\today}

\begin{abstract}
\noindent
We calculate the   formation probability of primordial black holes  generated during the collapse at horizon re-entry of large fluctuations produced during  inflation, such as those ascribed to a period of ultra-slow-roll. 
We show that it interpolates between a Gaussian at small values of the average density contrast and a Cauchy probability distribution at large values.
The corresponding abundance of primordial black holes may be larger than the Gaussian one by several orders of magnitude. The mass function is also shifted towards larger masses.

\end{abstract}

\maketitle

\paragraph{Introduction.}
\noindent
The recent release of the compact binary coalescences observed by LIGO and Virgo during the first half of the third observing run \cite{LIGO-Virgo}  has increased the interest  in the physics of Primordial Black Holes (PBHs) \cite{rev}. Indeed, there is a favourable  evidence  not only for multiple formation channels  in the BH mergers detected so far through the gravitational waves, but also for a  possible population of PBHs of about $30\%$ \cite{us}.
 
One of the basic parameters determining the merger rate of PBH binaries is  $f_{\textrm{\tiny PBH}}(M)$, the fraction  of PBHs with a given mass $M$ in the whole dark matter budget. It is roughly given by the expression   \cite{rev}
 \be
 \label{beta}
 f_{\textrm{\tiny PBH}}(M)\simeq
 \left(\frac{\beta(M)}{6\cdot 10^{-9}}\right)\left(\frac{M_\odot}{M}\right)^{1/2}.
  \ee
Here, the parameter $\beta(M)$ indicates the probability of formation of a PBH with a given mass $M$. It is clear therefore that a key question when dealing with the physics of PBHs is what is the formation probability of a PBH during the evolution of the universe. The answer depends, of course, on the details of the formation mechanism. 

A  standard  way  to generate    PBHs is  to enhance   the curvature perturbation $\zeta$   at small scales   \cite{s1,s2,s3, Kawasaki:1997ju}. Such boost   can occur either within single-field models of inflation during an ultra-slow-roll phase \cite{sasaki, Yokoyama:1998pt} or through some spectator field  \cite{curv}.  
The  enhancement needed for  the  power spectrum of the curvature perturbation  is from its $\sim 10^{-9}$ value at large scales to $\sim 10^{-2}$ on small scales. Such sizeable    fluctuations  are subsequently    communicated to radiation during the reheating process after inflation and they give rise  to PBHs upon horizon re-entry if they are larger than a given threshold.

From the brief discussion above it is clear that one main problem in determining the PBH abundance today is to deal with rare large fluctuations. The PBH formation probability  is extremely sensitive to changes in the tail of the fluctuation distribution and therefore to possible non-Gaussianities, primordial or intrinsic to the problem \cite{Yoo:2018kvb, byrnes,ine, Atal:2018neu, Kuhnel:2019xes}. The goal of this paper is to  show  that it is possible to deal with the non-linearities of the problem and obtain the   formation probability of PBHs. 

\vskip 0.3cm
\noindent
\paragraph{Setting the stage.} 
\noindent
Before launching ourselves in the midst of the discussions, let us set the stage and define the quantity we have to compute the probability of. The key parameter is the smoothed density contrast at the horizon crossing during radiation \cite{musco} 
\be
\delta_m=\frac{3}{\left(r_m e^{\zeta(r_m)}\right)^3}\int_0^{r_m} {\rm d} r\,\delta(r)\left(r e^{\zeta(r)}\right)^2\left(r e^{\zeta(r)}\right)',
\ee
 where $r_m$ is the location of the maximum of  the compaction function, which measures the mass excess compared to the background value in a given radius, and~\cite{Harada:2015yda}
 \be
 \delta(r)=-\frac{8}{9}\left(\frac{1}{aH}\right)^2 e^{-5\zeta(r)/2}\nabla^2 e^{\zeta(r)/2}
 \ee
 is the non-linear density contrast, $a$ is the scale factor and  $H$ the Hubble rate. 
 We have adopted the top-hat window function in order to correctly account for the treatment of the threshold~\cite{Young:2019osy}.
 This results in the non-linear expression \cite{musco,byrnes}
 \be
 \delta_m=\delta_l-\frac{3}{8}\delta_l^2,\,\,\,\, \delta_l=-\frac{4}{3}r_m\zeta'(r_m).
 \ee
It highlights two  key points. First of all, the probability of forming PBHs does not depend on the comoving curvature perturbation itself $\zeta$, but on its derivative $\zeta'$. This is expected, given that on superhorizon scales one can always add or subtract to the comoving curvature perturbation a constant by a coordinate transformation and  this  may not influence any physical result. Secondly, thanks to the conservation of the probability
\be
P(\delta_l)=P[\delta_m(\delta_l)]  \left|\frac{ {\rm d} \delta_m}{ {\rm d} \delta_l}\right|,
\ee
we ultimately need to compute the probability of $\delta_l$. Integrating it from the critical amplitude for $\delta_l$
\be
\delta_{l,c}=\frac{4}{3}\left(1-\sqrt{1-\frac{3}{2}\delta_c}\right),
\ee
in terms of the critical amplitude $\delta_c\simeq 0.59$ of $\delta_m$ (for a monochromatic curvature perturbation power spectrum) \cite{musco,mus}, one finds the probability of PBH formation.  The rest of the paper is dedicated
to the calculation of the   probability of $\delta_l$ in the standard ultra-slow-roll scenario.

\vskip 0.3cm
\noindent
\paragraph{The ultra-slow-roll scenario.}
\noindent
The ultra-slow-roll scenario (sometimes dubbed  also non-attractor scenario) of single-field models of inflation is a simple mechanism to enhance the curvature perturbation during inflation. It relies on the assumption that during its evolution the inflaton field $\phi$ traverses for a long enough period a plateau of its potential $V(\phi)$. If so, its equation of motion is approximately (primes here denote differentiation with respect to the number of e-folds $N$)
\be
\phi''(N)+3 \phi'(N)=0, \,\, \frac{{\rm d} V}{{\rm d}\phi}\simeq 0.
\ee
The solution is simply 
\begin{equation}
\phi(N)=\phi_e+\frac{\pi_e}{3}\left(1-e^{-3N}\right), \,\, \pi(N)\equiv \phi'(N)=\pi_e e^{-3N},
\end{equation}
where $\phi_e$ is the  value of the field at the end of the ultra-slow-roll phase.  We have retained the dependence on
$\pi$ since slow-roll is badly violated and also defined $N=0$ to be the end of the ultra-slow-roll phase, so that $N<0$. 

As we mentioned already, PBHs are born from  large, and therefore rare,   curvature perturbations. As such, their abundance is extremely
sensitive to the non-linearities of the curvature perturbation. A  particularly useful formalism when dealing with non-linearities is the so-called 
$\delta N$ formalism \cite{deltaN}, where  the   scalar field fluctuations are quantised on the flat slices and  $\zeta=-\delta N$, being $N$ the number of e-folds.
The formalism is based on the  assumption that, on superhorizon scales, each spatial point of the universe has an independent evolution and the latter is well approximated by the evolution of an unperturbed universe.

Using the $\delta N$ formalism we immediately find (see, for example, \cite{sasaki,Biagetti:2018pjj})
\be
\zeta=-\delta N=-\frac{1}{3}\ln\left(1+\frac{\delta\pi_e}{\overline{\pi}_e}\right),
\ee
where the overlines indicate the corresponding background values. By using the relation
\be
\pi_e=3\left[\phi(N)-\phi_e\right]+\pi(N),
\ee
we see that,  up to irrelevant constants,
\be
\label{zeta}
\zeta=-\frac{1}{3}\ln\left(1+3\frac{\delta\phi}{\overline{\pi}_e}\right).
\ee
This non-linear expression highlights the enhancement of the curvature perturbation due to the fact that the velocity
of the inflaton field becomes exponentially small during the ultra-slow-roll phase.

Since the dynamics of $\delta\phi$ is the one of a massless perturbation in de Sitter, and to a very good approximation its behaviour is Gaussian, 
the curvature perturbation is non-Gaussian due to the non-linear mapping between $\delta\phi$ and $\zeta$.
Probability conservation dictates that 
\begin{eqnarray}
P(\zeta)=P[\delta\phi(\zeta)]\left|\frac{{\rm d}\delta\phi}{{\rm d} \zeta}\right|.
\end{eqnarray}
Since $P(\delta\phi)$ is Gaussian
\begin{eqnarray}
P(\delta\phi)=\frac{1}{\sqrt{2\pi}\,\sigma_{\delta\phi}}e^{-(\delta\phi)^2/2\sigma_{\delta\phi}^2},
\end{eqnarray}
where $\sigma_{\delta\phi}^2$ is the variance of the inflaton fluctuations
\begin{eqnarray}
\sigma_{\delta\phi}^2=\int {\rm d}\ln k \,{\cal P}_{\delta\phi}(k)
\end{eqnarray}
in terms of the inflaton perturbation power spectrum ${\cal P}_{\delta\phi}(k)$, in the limit of small fluctuations $\zeta\ll 1$ one recovers a Gaussian distribution for the
curvature perturbation. However, for large values of the curvature perturbation one gets
\begin{eqnarray}
P(\zeta)\simeq \frac{\overline{\pi}_e}{\sqrt{2\pi}\,\sigma_{\delta\phi}} e^{-3\zeta},
\end{eqnarray}
showing that the probability of the curvature perturbation is of the exponential type and has a  non-Gaussian exponential  tail. This result confirms what found in Refs. \cite{d1,d2,d3} using a stochastic approach,  in the realistic limit of large field displacement during the ultra-slow roll phase in units of the Hubble rate.  It is also valid in the limit in which the subsequent transition into the slow-roll phase is rapid~\cite{sasaki} and for a constant potential (for a slightly tilted potential see the results of Refs.~\cite{sasaki, Atal:2019cdz, Atal:2019erb}).
However, as mentioned in the introduction, this is not the end of the story, one needs to in fact calculate the probability of the density contrast $\delta_l$ and not of the curvature perturbation $\zeta$ to compute the abundance of PBHs. 

We close this section by noticing that the same non-linear relation between the curvature perturbation and the inflaton fluctuation in Eq.~(\ref{zeta}) is also obtained if the ultra-slow-roll phase is anticipated by a short period during which the inflaton  accelerates by falling down a rapid step in the potential \cite{fall}. This is because the contribution to the curvature perturbation from this phase using the $\delta N$ formalism is suppressed. Similarly, the expression (\ref{zeta}) is also obtained in the case in which the PBHs are formed thanks to the large fluctuations in a curvaton-like field \cite{curv}, where the fluctuation of the inflaton field is replaced by the one in the curvaton-like field \cite{ws}. In this respect, our considerations are rather general and go beyond the single-field ultra-slow-roll mechanism.

\vskip 0.3cm
\noindent
\paragraph{The formation probability of PBHs.}
\noindent
As already remarked, the next step is to compute the probability distribution for the field 
\be
\delta_l=-\frac{4}{3}r_m\zeta'(r_m)=\frac{4}{3}r_m\frac{\delta\phi'(r_m)/\overline{\pi}_e}{1+3\,\delta\phi(r_m)/\overline{\pi}_e}.
\ee
One can do it exactly as $\delta_l$ is the ratio of two normally distributed and uncorrelated  random variables 
\begin{eqnarray}
X&=&\frac{4}{3} r_m \frac{\delta\phi'(r_m)}{\overline{\pi}_e},\nonumber\\
Y&=&1+3\frac{\delta\phi(r_m)}{\overline{\pi}_e},
\end{eqnarray}
and therefore its probability is the ratio distribution
\begin{eqnarray}
P(Z)\equiv P(X/Y)&=&\int{\rm d}Y'\, |Y'| P(ZY', Y'),\nonumber\\
P(X,Y)&=&\frac{1}{2\pi\sqrt{\det C}}\exp\left(-\vec{V}^T C^{-1}\vec V/2\right),\nonumber\\
\vec{V}^T&=&(X,Y-1), \ C={\rm diag}\left(
\sigma_{X}^2,\sigma_Y^2\right), \nonumber \\
&&
\end{eqnarray}
where
\begin{eqnarray}
\sigma_X^2=\frac{16r_m^2}{9\overline{\pi}_e^2}\int {\rm d}\ln k \,k^2 \,{\cal P}_{\delta\phi}(k),
\
\sigma_Y^2=\frac{9}{\overline{\pi}_e^2} \sigma^2_{\delta\phi},
\end{eqnarray}
and the cross correlation between $X$ and $Y$ is zero because of isotropy.
We then obtain
\begin{eqnarray}
\label{full}
&&P(\delta_l)=\frac{e^{-1/2\sigma_Y^2}}{2\pi(\sigma_X^2+\sigma_Y^2 \delta_l^2)^{3/2}}\sigma_X\cdot\left[2\sigma_Y\sqrt{\sigma_X^2+\sigma_Y^2 \delta_l^2}\right.\nonumber\\
&+&\left. \sqrt{2\pi}\sigma_X e^{\sigma_X^2/(2\sigma_X^2\sigma_Y^2+2\sigma_Y^4 \delta_l^2)}{\rm Erf}\left(\frac{\sigma_X}{\sqrt{2}\sigma_Y\sqrt{\sigma_X^2+\sigma_Y^2 \delta_l^2}}\right)\right].\nonumber\\
&&
\end{eqnarray}
In the limit of small inflaton perturbations, namely  $\sigma_Y\delta_l\ll 1$, this expression reduces, as it should,  to a simple Gaussian distribution,
\be
\label{gaussian}
P(\delta_l)\simeq \frac{e^{-\delta_l^2/2\sigma_X^2}}{\sqrt{2\pi}\sigma_X}.
\ee
In the limit of large density contrast fluctuations $\sigma_Y\delta_l\gg 1$, we get 
\be
P(\delta_l)\simeq e^{-1/2\sigma_Y^2}\frac{\sigma_X}{\pi\sigma_Y \delta_l^2},
\ee
which is the   Cauchy probability distribution for large values of the field. 

The  full probability is neither a Gaussian nor an  exponential distribution. This conclusion, as well as the expression (\ref{full}), is the main results of this paper. 

The last step  to compute the probability of forming a PBH is to calculate the mass fraction (taking into account that $\delta_m<2/3$ \cite{musco})
\be\label{mass fraction}
\beta(M_H)=\int_{\delta_{l,c}}^{4/3}{\rm d}\delta_l \frac{M}{M_H} P(\delta_l),
\ee
where we keep into account the relation between the horizon mass $M_H$ at formation and the PBH mass for overdensities close to the critical threshold as~\cite{Choptuik:1992jv, Niemeyer:1997mt, Evans:1994pj, byrnes}
\be\label{masscriticalcoll}
M = \kappa M_H \llp \lp \delta_l - \frac{3}{8} \delta_l^2 \rp - \delta_{c}\rrp^\gamma,
\ee
where $\kappa=3.3$ and $\gamma=0.36$ for the collapse in a radiation-dominated universe.
 \begin{figure}[t!]
 	\centering
 	\includegraphics[width=1 \linewidth]{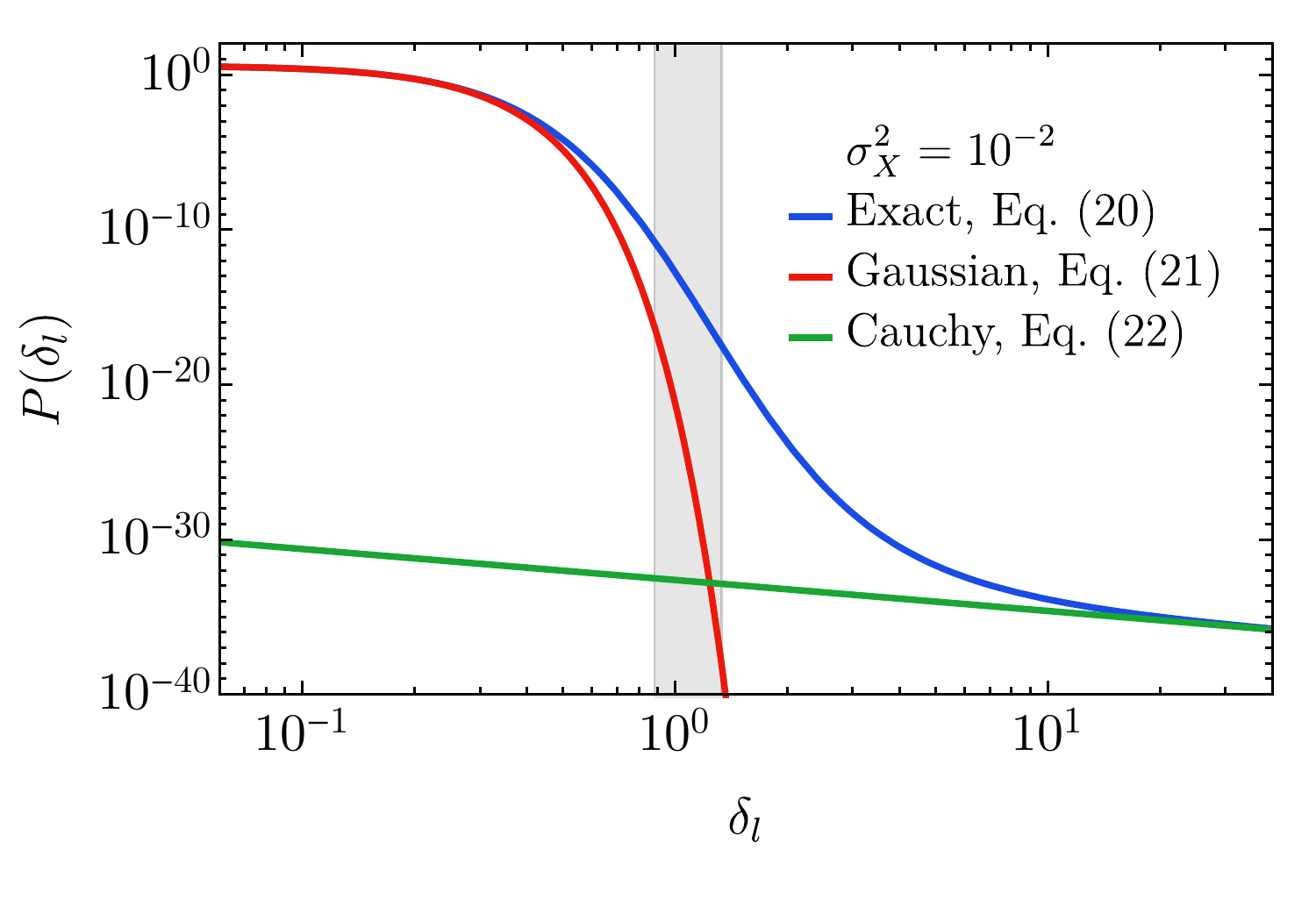}
 	\caption{\it The probability of the density contrast for fixed variance $\sigma_X^2$ and a monochromatic power spectrum of inflaton fluctuations. The blue line denotes the result of Eq.~\eqref{full}, while the red and green lines indicate the limit for small and large linear density perturbations $\delta_l$, respectively. The vertical band indicates the relevant range of values of the linear density contrast for the PBH formation probability.}
 	\label{fig: 1}
 \end{figure}
 \begin{figure}[t!]
	\centering
	\includegraphics[width=1 \linewidth]{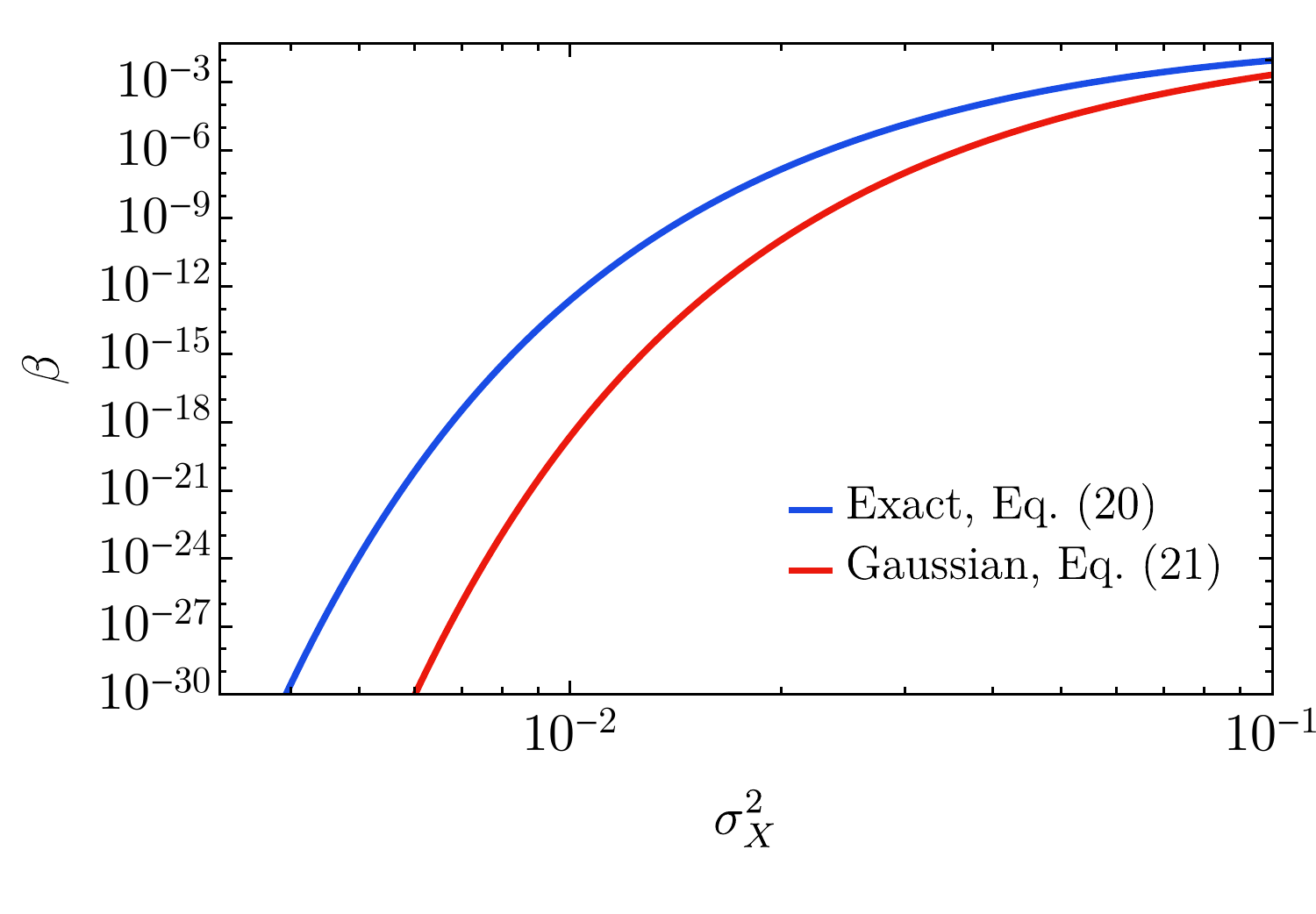}
	\caption{\it The PBH mass fraction, as defined in Eq.~\eqref{mass fraction}, for a narrow spectrum of the inflaton  perturbations.}
	\label{fig: 2}
\end{figure}
 From Eq.~\eqref{beta} we see that $f_{\textrm{\tiny PBH}}(M)\lsim 1$ corresponds to  tiny  fractions,  $\beta(M)\lsim 6\cdot 10^{-9}(M/M_\odot)^{1/2}$. For such values, the PBH abundance is sensitive to the non-Gaussian tail of the distribution.
 In Fig. 1 we plot the probability (\ref{full}) for a representative monochromatic power spectrum of the inflaton fluctuations ${\cal P}_{\delta\phi}(k)=A k_\star \delta(k-k_\star)$
peaked at a momentum scale $k_\star$ such that $r_m k_\star\simeq 3$ \cite{musco}, for which
$\sigma_Y^2\simeq (9/16) \sigma_X^2$. This figure highlights that, in the relevant range of values of $\delta_l$, the probability is neither a Gaussian nor a Cauchy distribution. 

In Fig.~2 we plot the PBH  fraction as a function of $\sigma_X^2$, showing that
the final abundance is enhanced compared to the Gaussian case by various orders of magnitude. 

In Fig.~3 we plot the PBH mass function, showing that it has more power at high masses than the Gaussian prediction. This difference is due to 
a weaker suppression of $\delta_m$ entering in the critical collapse relation \eqref{masscriticalcoll} when the probability distribution \eqref{full} is adopted. 

\begin{figure}[t!]
	\centering
	\includegraphics[width=1 \linewidth]{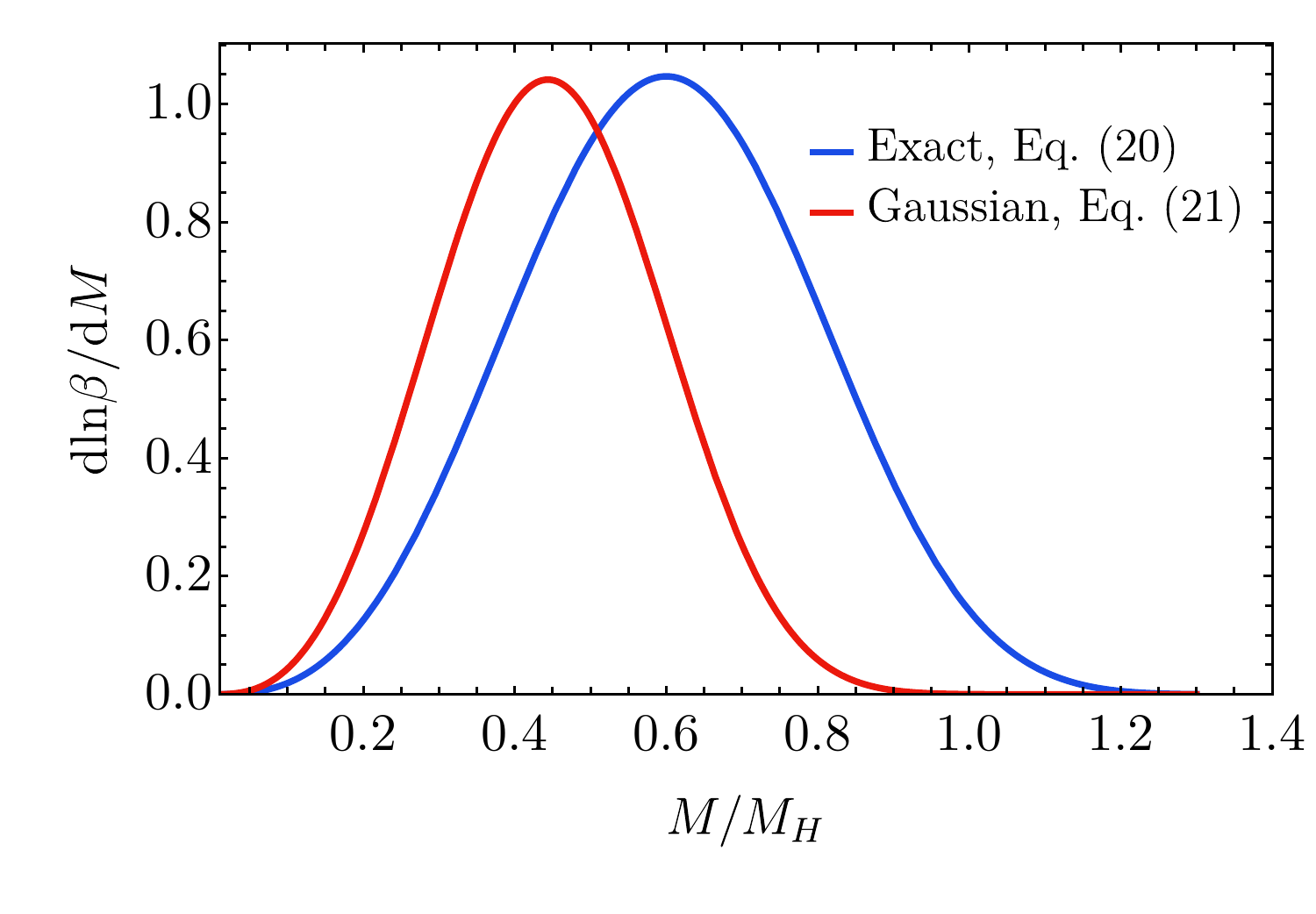}
	\caption{\it The PBH mass function for the same assumptions of Fig. 2.}
	\label{fig: 3}
\end{figure}

\vskip 0.3cm
\noindent
\paragraph{Conclusions.}
\noindent
PBHs might play a significant role in the BH mergers seen through the gravitational waves by the LIGO-Virgo collaboration. They might also contribute significantly to the dark matter of the universe if their mass is in the asteroid range \cite{bartolo}. In this paper we have provided the probability of formation in the scenarios in which PBHs are formed thanks to the collapse, upon horizon re-entry, of large perturbations generated during inflation. Our findings indicate that such a probability is    non-Gaussian at large values of the average density contrast, which is the correct variable to use when computing the PBH abundance \cite{Young:2014ana}.   Furthermore, our results show that
not only the corresponding PBH abundance is larger than the Gaussian result by orders of magnitude, but also the mass function has a more pronounced tail at larger masses.

\vskip 0.3cm
\noindent
\paragraph{Acknowledgments.}
\noindent
We thank V. Atal, J. Garriga and D. Wands for interesting discussions.
M.B.  acknowledges  support  from  the  Netherlands Organization for Scientific Research (NWO), which is funded by the Dutch Ministry of Education, Culture and Science (OCW), under VENI grant 016.Veni.192.210. V.DL., G.F. and 
A.R. are supported by the Swiss National Science Foundation 
(SNSF), project {\sl The Non-Gaussian Universe and Cosmological Symmetries}, project number: 200020-178787.

\end{document}